# QCD vacuum tensor susceptibility and properties of transversely polarized mesons

A. P. Bakulev and S. V. Mikhailov

*Bogoliubov Theoretical Laboratory, Joint Institute for Nuclear Research, 141980, Dubna, Russia*

(September 23, 2018)

We re-estimate the tensor susceptibility of QCD vacuum, $\chi$, and to this end, we re-estimate the leptonic decay constants for transversely polarized $\rho$- , $\rho'$- and $b_1$-mesons. The origin of the susceptibility is analyzed using duality between $\rho$- and $b_1$- channels in a 2-point correlator of tensor currents. We confirm the results of [1] for the 2-point correlator of tensor currents and disagree with [2] on both OPE expansion and the value of QCD vacuum tensor susceptibility. Using our value for the latter we determine new estimations of nucleon tensor charges related to the first moment of the transverse structure function $h_1$ of a nucleon.

## INTRODUCTION

In this paper, we investigate the low-energy properties of the lightest transversely polarized mesons with quantum numbers $J^{PC} = 1^{--}(\rho)$, $1^{+-}(b_1)$ in the framework of QCD sum rules (SRs) with nonlocal condensates (NLCs) as well as with the standard ones. This work was started in [3] where the "mixed parity" NLC SR for the light-cone distribution (LCD) amplitudes of both $\rho$- and $b_1$-mesons was constructed. It was concluded that to obtain a reliable result we should reduce model uncertainties due to the nonlocal gluon contribution into the SR for LCD. Different SRs for each P-parity could be preferable for this purpose. As a first step, to obtain twist 2 meson LCD, we concentrate on the meson static properties using the "pure parity" NLC SR for each meson separately:

1) we re-estimate the leptonic decay constants $f_m^T$ for transversely polarized $\rho(770)$, $\rho'(1465)$-mesons $(1^{--})$ and the $b_1(1235)$-meson $(1^{+-})$ [4];

2) we correct the previous consideration by Belyaev and Oganesyan (B&O) [2] and provide a new estimation for the vacuum tensor susceptibility (VTS) introduced in [5,6].

The static characteristics, the decay constants $f_m^T$ and "continuum thresholds" $s_m$ (parameters of phenomenological models for spectral densities) of the lightest transversely polarized mesons in the channels with $J^{PC} = 1^{--}$ and $1^{+-}$, are tightly connected with the value of VTS. Namely, the difference of the meson properties in these channels fixes the non-zero value of VTS: in a hypothetical Nature, *e.g.*, where the properties of these mesons are the same, VTS is identically equal to zero. For the reason that these meson constants should appear in VTS in a form of a difference, we have to define them more precisely and in the framework of a unified approach.

The method of NLC SRs was successfully applied for the determination of meson dynamic characteristics (LCD amplitudes, form factors, see, *e.g.*, [3,7] and refs therein). One of its basic components is a non-zero characteristic scale, $\lambda_q$, of quark-gluon correlations in the QCD vacuum [7]. This parameter fixes the average virtuality of vacuum quarks which flow through vacuum with the momentum $k_q$, $\langle k_q^2 \rangle = \lambda_q^2 \approx 0.4 - 0.5$ GeV$^2$ [10], this value is of an order of hadronic scale, $m_\rho^2 \approx 0.6$ GeV$^2$, and is of importance in the calculations. Note here that quite recently the nonlocal character of the quark condensate has been confirmed in the lattice calculations [11], where an attempt to estimate $\lambda_q$ was made. NLC approach can improve the stability and accuracy of SRs even for the case of decay constant determination where the NLC effect is of an order of the radiative correction contribution. Therefore, we revise their values in pure parity NLC SRs, despite the presence of different estimations for these quantities in literature [8,2,3], obtained by different ways. For comparison we also calculate all these quantities by the standard way, that corresponds to processing our NLC SR on the limit $\lambda_q^2 \to 0$.

The source of the above-mentioned difference of meson properties is the peculiarity of four-quark condensate contribution to the "theoretical part" of SRs. This contribution is invariant under the duality[1] transformation in contrast to all other condensate contributions which change the sign under the same transformation. This peculiarity of four-quark condensate contribution will be considered in detail.

The plan of presentation is the following: firstly, we discuss the QCD SR approach to investigation of the 4-rank tensor 2-point correlator for transversely polarized $\rho$-, $\rho'$- and $b_1$-mesons. Then, we define the duality transformation and draw its consequences for the constructed SRs. Finally, we derive a new estimation for the QCD VTS and nucleon tensor charges and discuss what is wrong in the consideration of [2].

### DECAY CONSTANTS OF TRANSVERSELY POLARIZED ($J^{PC} = 1^{--}, 1^{+-}$)-MESONS

We start with a 2-point correlator of tensor currents $J^{\mu\nu}(x) = \bar{u}(x)\sigma^{\mu\nu}d(x)$,

$$\Pi^{\mu\nu;\alpha\beta}(q) = \int d^4x \; e^{iq\cdot x} \langle 0|T[J^{\mu\nu+}(x)J^{\alpha\beta}(0)]|0\rangle. \tag{1}$$

(Note here that due to isospin symmetry this is the same correlator which was studied in [2].) This correlator can be decomposed in invariant form factors $\Pi_\pm$, [1,8]

$$\Pi^{\mu\nu;\alpha\beta}(q) = \Pi_-(q^2)P_1^{\mu\nu;\alpha\beta} + \Pi_+(q^2)P_2^{\mu\nu;\alpha\beta} \tag{2}$$

where the projectors $P_{1,2}$ are defined by the expressions

$$P_1^{\mu\nu;\alpha\beta} \equiv \frac{1}{2q^2}\left[g^{\mu\alpha}q^\nu q^\beta - g^{\nu\alpha}q^\mu q^\beta - g^{\mu\beta}q^\nu q^\alpha + g^{\nu\beta}q^\mu q^\alpha\right] \; ; \tag{3}$$

$$P_2^{\mu\nu;\alpha\beta} \equiv \frac{1}{2}\left[g^{\mu\alpha}g^{\nu\beta} - g^{\mu\beta}g^{\nu\alpha}\right] - P_1^{\mu\nu;\alpha\beta} \; , \tag{4}$$

which obey the projector-type relations

$$(P_i \cdot P_j)^{\mu\nu;\alpha\beta} \equiv P_i^{\mu\nu;\sigma\tau}P_j^{\sigma\tau;\alpha\beta} = \delta_{ij}P_i^{\mu\nu;\alpha\beta} \text{ (no sum over } i\text{)}, \; P_i^{\mu\nu;\mu\nu} = 3. \tag{5}$$

Then, for the form factors $\Pi_\pm(q^2)$ it is possible to use dispersion representations of the form

$$\Pi_\pm(q^2) = \frac{1}{\pi}\int_0^\infty \frac{\rho_\pm(s)ds}{s - q^2} + \text{subtractions} \; , \tag{6}$$

which after the Borel transformation (with Borel parameter $M^2$) become

---

[1] We are grateful to O. V. Teryaev who involved us in an investigation of this transformation and suggested this name for it.

$$\Pi_\pm(q^2) \to B\Pi_\pm(M^2) = \frac{1}{\pi M^2} \int_0^\infty \rho_\pm(s) e^{-s/M^2} ds \ . \tag{7}$$

A phenomenological model for the spectral density $\rho^{\text{phen}}(s)$ is usually taken in the form of "lowest resonances + continuum"

$$\rho_\pm^{\text{phen}}(s) = \pm 2\pi \left|f_m^T\right|^2 s \cdot \delta(s - m_m^2) + \rho_\pm^{\text{pert}}(s)\theta(s - s_\pm) \ , \tag{8}$$

where $f_m^T$ and $m_m$ are the decay constants and masses of the lowest meson resonances, $m = \rho, \rho', b_1$, contributing to the correlator of interest, and $\rho_\pm^{\text{pert}}(s)$ are the corresponding spectral densities of perturbative contributions to the correlators $\Pi_\pm(q^2)$. The decay constants are defined via the parameterization of the unit helicity ($|\lambda| = 1$) states of $\rho$-, $\rho'$- and $b_1$-mesons

$$\langle 0 \left|\bar{u}(x)\sigma_{\mu\nu}d(x)\right| \rho^+(p, \lambda)(\rho'^+)\rangle = if_{\rho,\rho'}^T \left(\varepsilon_\mu(p, \lambda)p_\nu - \varepsilon_\nu(p, \lambda)p_\mu\right) \ ; \tag{9}$$

$$\langle 0 \left|\bar{u}(x)\sigma_{\mu\nu}d(x)\right| b_1^+(p, \lambda)\rangle = f_{b_1}^T \epsilon_{\mu\nu\alpha\beta}\varepsilon^\alpha(p, \lambda)p_\beta \ , \tag{10}$$

here $\varepsilon^\mu(p, \lambda)$ is the polarization vector of a meson with momentum $p$ and helicity $\lambda$. To construct SRs, one should calculate OPE of the correlators $\Pi_\pm(M^2)$

$$B\Pi_\pm(M^2) = \frac{1}{\pi M^2} \int_0^\infty \rho_\pm^{\text{pert}}(s) e^{-s/M^2} ds + \frac{a_\pm}{M^2}\langle\frac{\alpha_s}{\pi}G^2\rangle + \frac{b_\pm}{M^4}\pi\langle\sqrt{\alpha_s}\bar{q}q\rangle^2 \ . \tag{11}$$

We perform these calculations in the approach of QCD SRs with NLCs (see [3]), where the coefficients $a_\pm, b_\pm$ become functions $a_\pm(M^2), b_\pm(M^2)$ of the Borel parameter $M^2$ which tend to their standard values for large $M^2$, $M^2 \gg \lambda_q^2$, e.g. , $b_\pm = \lim_{(\lambda_q^2/M^2) \to 0} b_\pm(M^2)$. The functions $a_\pm(M^2), b_\pm(M^2)$ can be considered as accumulating an infinite subset of the standard condensate $\left(\lambda_q^2/M^2\right)^j$-contributions [7] in OPE. All needed NLC expressions are given in Appendix A, while the standard coefficients $a_\pm, b_\pm$, corresponding to the limit $\lambda_q^2/M^2 \to 0$, are explicitly written below. Their values are in full agreement with the preceding calculations performed in [1] and [8]

$$\frac{1}{(\pm 2)}\rho_\pm^{\text{pert}}(s) = \rho_0^{\text{pert}}(s) \equiv \frac{s}{8\pi}\left[1 + \frac{\alpha_s(\mu^2)}{\pi}\left(\frac{7}{9} + \frac{2}{3}\log\frac{s}{\mu^2}\right)\right] \ ; \tag{12}$$

$$\frac{1}{(\pm 2)}a_\pm = \frac{1}{24} \ ; \tag{13}$$

$$\frac{1}{(-2)}b_- = \frac{-16 + 80 + 144}{81} = \frac{208}{81} \tag{14}$$

$$\frac{1}{(+2)}b_+ = \frac{-16 + 80 - 144}{81} = \frac{-80}{81} \ . \tag{15}$$

Here $\mu$ is the renormalization scale ($\mu^2 \simeq 1 \text{ GeV}^2$) and the coefficients listed in the central parts of the last two lines correspond to the vector $\langle\bar{q}\gamma_m q\rangle$, quark-gluon-quark $\langle\bar{q}G_{\mu\nu}q\rangle$ and the four-quark $\langle\bar{q}q\bar{q}q\rangle$ vacuum condensate contributions (see details in Appendix A, [7]). We write down these coefficients explicitly in order to reveal the discrepancy between our results and those obtained by B&O [2], who found, instead, in the last line

$$\frac{-16 - 48 - 144}{81} = \frac{-208}{81},$$

a result larger than ours by a factor of 2.6. We conclude that in [2] there is a wrong contribution due to the quark-gluon-quark vacuum condensate.

Collecting all parts (7), (8), (11) together, one obtains the following SRs:

$$|f_\rho^T|^2 m_\rho^2 e^{-m_\rho^2/M^2} + (\rho \to \rho') = \frac{1}{\pi}\int_0^{s_\rho} \rho_0^{\text{pert}}(s) e^{-s/M^2} ds - \frac{a_-}{2}\langle\frac{\alpha_s}{\pi}G^2\rangle - \frac{b_-(M^2)}{2M^2}\pi\langle\sqrt{\alpha_s}\bar{q}q\rangle^2 \; ; \quad (16)$$

$$|f_{b_1}^T|^2 m_{b_1}^2 e^{-m_{b_1}^2/M^2} = \frac{1}{\pi}\int_0^{s_{b_1}} \rho_0^{\text{pert}}(s) e^{-s/M^2} ds + \frac{a_+}{2}\langle\frac{\alpha_s}{\pi}G^2\rangle + \frac{b_+(M^2)}{2M^2}\pi\langle\sqrt{\alpha_s}\bar{q}q\rangle^2. \quad (17)$$

The role of NLC, concentrated in $a_\pm, b_\pm(M^2)$, is important here, *i.e.*, at $M^2 = 0.6$ GeV$^2$ the total condensate contribution in the SR reduces twice in comparison with the standard (local) one. In accordance with QCD SR practice the processing of these NLC SRs are performed within the validity window $M_-^2 \leq M^2 \leq M_+^2$ (see details in [9,3]). These windows are determined by two conditions: the lower bound $M_-^2$ by demanding that the relative value of $\langle GG \rangle$- and $\langle \bar{q}q \rangle$-contributions to OPE series should not be larger than 30%, the upper one $M_+^2$ by requiring that a relative contribution of higher states in the phenomenological part of SR should not be larger than 30%. The processing with the standard values of vacuum condensates (see Appendix A) gives the decay constants

$$f_\rho^T = 0.157 \pm 0.005 \text{ GeV}, \quad f_{\rho'}^T = 0.140 \pm 0.005 \text{ GeV}, \quad s_{\rho,\rho'}^T = 2.8 \text{ GeV}^2 \; ; \quad (18)$$

$$f_{b_1}^T = 0.184 \pm 0.005 \text{ GeV}, \quad s_{b_1}^T = 2.87 \text{ GeV}^2 , \quad (19)$$

which are presented at normalization point $\mu^2 = 1$ GeV$^2$. Very stable curves in wide validity windows have been obtained for all of these quantities.

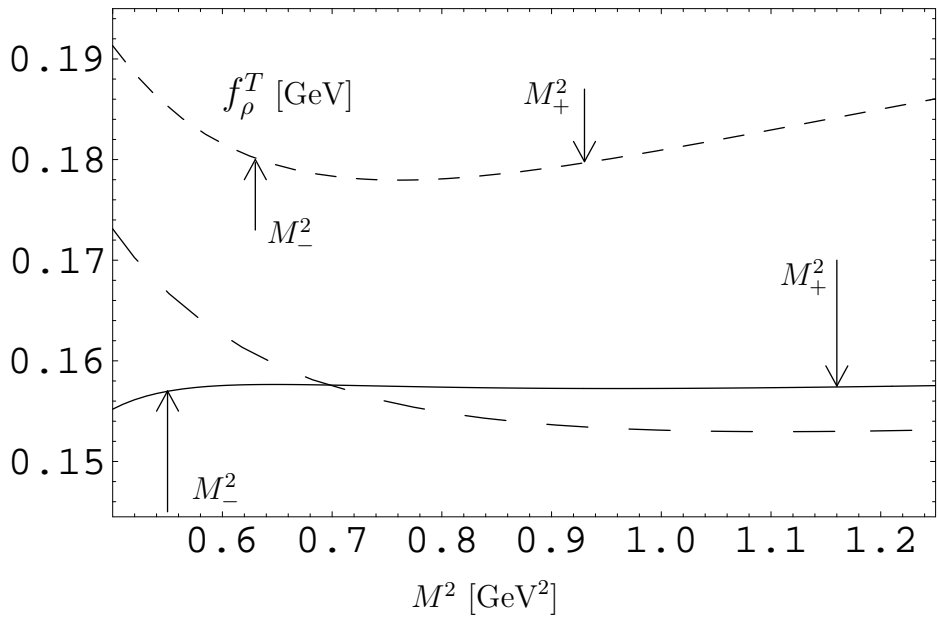

FIG. 1. The curves of $f_\rho^T$ in $M^2$; the solid line corresponds to NLC SR with the $\rho'$-meson taken into account, the long arrows show its validity window; the short-dashed line corresponds to the standard SR without $\rho'$-meson, the small arrows show reduced validity window for this case; the dashed line corresponds to B&B analyses.

The processing of the "local" version (at $\lambda_q^2 \to 0$) of the SRs (16)-(17) leads to the values [2]:

---

[2]To provide a clear comparison with the results of B&O, who do not take into account $\rho'$-meson contribution, condensate

$$f_\rho^T = 0.179(0.170) \pm 0.007 \text{ GeV}, \quad f_{\rho'}^T \sim 0 \text{ GeV}, \quad s_{\rho,\rho'}^T = 2.1 \text{ GeV}^2 \;; \tag{20}$$

$$f_{b_1}^T = 0.191(0.178) \pm 0.009 \text{ GeV}, \quad s_{b_1}^T = 3.2 \text{ GeV}^2 \;, \tag{21}$$

which accuracy looks worse. Really, the curve corresponding to $f_\rho^T$ in $M^2$ is shown in Fig. 1 (solid line) in comparison with the result of the standard approach without $\rho'$-meson (short-dashed line). For the first case, the validity window expands in all the region $(0.55 - 1.20)$ GeV$^2$ (long arrows) while for the latter case it shrinks twice to the region denoted in figure by the small arrows $M_-^2$ and $M_+^2$. Note, that the standard SR "pushes out" the $\rho'$ meson and does not allow to obtain his parameters, while the NLC SR is sensitive to this meson and even allows to determine its mass [3]. We demonstrate on the same figure the curve for $f_\rho^T$ (dashed line) obtained in [8] by Ball and Braun (B&B) in the framework of the standard approach, the same small arrows denoting its real "working window". Note here that processing B&B SR just in this thin working window results in a curve very similar in shape to the upper short-dashed one with the average value $f_\rho^T(1\text{GeV}^2) = 0.171$ GeV. Note, that recently performed lattice estimates [12] [3] give $f_{\rho Latt}^T(4\text{GeV}^2) = 0.165(11)$ GeV, that approximately agree with both the NLC (18) and the "standard" (20) values. So, we can conclude that our improved SRs (16)–(17) are really justified and produce reliable and stable results. All the results obtained by processing "pure parity" (16, 17) and "mixed parity" NLC SR [3] are collected in Table 1, in comparison with the previous results in [8,2].

|  | "pure parity" SR based on $\Pi_\mp$ (− for $\rho$, + for $b_1$) | | | "mixed parity" SR based on $(\Pi_- - \Pi_+)/q^2$ | |
|---|---|---|---|---|---|
| Source | Here | B&B [8] | B&O [2] | Here[4] | B&B [8] |
| $f_\rho^T$ [MeV] | 157(5) | 160(10) | − | 166(6) | 163(5) |
| $f_{\rho'}^T$ [MeV] | 140(5) | − | − | − | − |
| $s_\rho$ [GeV$^2$] | 2.8 | 1.5 | − | 1.5 | 2.1 |
| $f_{b_1}^T$ [MeV] | 184(5) | 180(10) | 178(10) | 179(7) | 180$^{\text{fixed}}$ |
| $s_{b_1}$ [GeV$^2$] | 2.87 | 2.7 | 3.0 | 2.93 | 2.1 |

**Table 1**. Estimates for decay constants $f^T(1\text{GeV}^2)$ of transversely polarized $\rho(770)$, $\rho'(1465)$- and $b_1(1235)$-mesons based on processing QCD SRs in different approaches.

---

nonlocality and $\alpha_s$-corrections in the perturbative spectral density, we write down the results of processing our SRs in the same approximation in parentheses.
[3]We are indepted to D. Becirevic, who informed us about these interesting papers, contaning lattice estimates mass and decay constants of mesons.
[4]The estimates presented in this column have been obtained by processing the "mixed parity" SR established in [3]. We improve the model for phenomenological spectral density using the features of phenomenological spectral densities of "pure parity" SRs.

It is interesting to note that in spite of the discrepancy in the OPE coefficients, the authors of [2] obtain for $f_{b_1}^T$ a value of $178 \pm 10$ MeV which is quite close to the value found by B&B [8]: $180 \pm 10$ MeV. This compensation effect happens due to the fact that both groups of authors used different sets of the condensate input-parameters in the SR and this resulted in approximately the same overall contributions of the quark condensate: B&B had $(\frac{1}{2}b_+)\pi\alpha_s\langle\bar{q}q\rangle^2 = -4.22\ 10^{-4}$ GeV$^6$; and B&O, $(\frac{1}{2}b_+)\pi\alpha_s\langle\bar{q}q\rangle^2 = -4.92\ 10^{-4}$GeV$^6$, see Appendix B.

## DUALITY AND ITS BREAKDOWN

Let us consider now an operator $\hat{D}$ transforming any rank-4 tensor $T^{\mu\nu;\alpha\beta}$ to another rank-4 tensor $T_D^{\mu\nu;\alpha\beta} = (\hat{D}T)^{\mu\nu;\alpha\beta}$ with

$$D^{\mu\nu;\alpha\beta}_{\mu'\nu';\alpha'\beta'} = \frac{-1}{4}\epsilon^{\mu\nu}{}_{\mu'\nu'}\epsilon_{\alpha'\beta'}{}^{\alpha\beta} \quad \text{and} \quad \hat{D}^2 = 1. \tag{22}$$

Our projectors $P_1^{\mu\nu;\alpha\beta}$ and $P_2^{\mu\nu;\alpha\beta}$ under the action of this operator transform into each other

$$\left(\hat{D}P_1\right)^{\mu\nu;\alpha\beta} = P_2^{\mu\nu;\alpha\beta}\ ;\quad \left(\hat{D}P_2\right)^{\mu\nu;\alpha\beta} = P_1^{\mu\nu;\alpha\beta}, \tag{23}$$

whereas the correlator $\Pi^{\mu\nu;\alpha\beta}(q)$ transforms into the correlator of dual tensor currents $J_5^{\mu\nu}(x) = \bar{u}(x)\sigma^{\mu\nu}\gamma_5 d(x)$

$$(\hat{D}\Pi)^{\mu\nu;\alpha\beta}(q) = \int d^4x\ e^{iq\cdot x}\langle 0|T[J_5^{\mu\nu+}(x)J_5^{\alpha\beta}(0)]|0\rangle\ . \tag{24}$$

There is a good question: how are $\Pi^{\mu\nu;\alpha\beta}(q)$ and $(\hat{D}\Pi)^{\mu\nu;\alpha\beta}(q)$ connected?

In perturbative QCD with massless fermions, taking into account the standard anticommutations, one easily arrives at

$$(\hat{D}\Pi)^{\mu\nu;\alpha\beta}_{\text{pert}}(q) = -\Pi^{\mu\nu;\alpha\beta}_{\text{pert}}(q), \tag{25}$$

from which it follows that $\Pi^{\mu\nu;\alpha\beta}_{\text{pert}}(q)$ is anti-self-dual. The same (anti-dual) character is inherent in the phenomenological models, see Eq. (8).

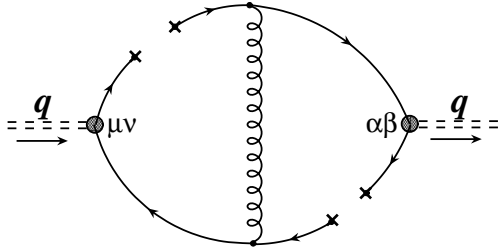

FIG. 2. Diagram with insertion of four-quark condensate.

The same reasoning is valid almost for all OPE diagrams, those with a gluon condensate, with a vector quark condensate, and with a quark-gluon-quark condensate. Only the diagram with four-quark scalar condensates is different (see Fig. 2): in that case there are 2 $\gamma$-matrices on one line between two external vertices (1 from the fermion propagator and 1 from the quark-gluon vertex) because the scalar condensate cancels one $\gamma$-matrix. Thus, we realize that the OPE contribution involves two parts, one being anti-self-dual (**ASD**) and the other one self-dual (**SD**)

$$\Pi^{\mu\nu;\alpha\beta}_{\text{OPE}}(q) = \mathbf{ASD}^{\mu\nu;\alpha\beta}(q) + \mathbf{SD}^{\mu\nu;\alpha\beta}(q) \;, \tag{26}$$

$$(-\hat{D}\mathbf{ASD})^{\mu\nu;\alpha\beta}(q) = \mathbf{ASD}^{\mu\nu;\alpha\beta}(q) \equiv \Pi^{\text{ASD}}(q^2)\left(P_1^{\mu\nu;\alpha\beta} - P_2^{\mu\nu;\alpha\beta}\right) \;, \tag{27}$$

$$(\hat{D}\mathbf{SD})^{\mu\nu;\alpha\beta}(q) = \mathbf{SD}^{\mu\nu;\alpha\beta}(q) \equiv \Pi^{\text{SD}}(q^2)\left(P_1^{\mu\nu;\alpha\beta} + P_2^{\mu\nu;\alpha\beta}\right) \;. \tag{28}$$

The appearance of the **SD**-diagrams breaks the anti-duality of the two correlators $\Pi$ and $\left(\hat{D}\Pi\right)$.

We can rewrite formulae (26)–(28) to obtain the following representation for the OPE-induced part of correlator:

$$\Pi^{\mu\nu;\alpha\beta}_{\text{OPE}}(q) = P_1^{\mu\nu;\alpha\beta}\left[\Pi^{\text{SD}}(q^2) + \Pi^{\text{ASD}}(q^2)\right] + P_2^{\mu\nu;\alpha\beta}\left[\Pi^{\text{SD}}(q^2) - \Pi^{\text{ASD}}(q^2)\right] \;. \tag{29}$$

As a simple consequence of this representation and Eq.(5) we have a useful relation

$$\Pi^{\mu\nu;\mu\nu}_{\text{OPE}}(q) = 6\,\Pi^{\text{SD}}(q^2) \;. \tag{30}$$

Using relations (27)–(28), one can easily calculate the OPE coefficients for different diagrams. For example, let us consider the $\langle\bar{q}Gq\rangle$-condensate and its contribution to the coefficient $b_\pm$. Indeed, we know that this contribution is of the **ASD**-type, that is

$$\Pi^{\mu\nu;\alpha\beta}_{\langle\bar{q}Gq\rangle} = c(q^2)\left(P_1^{\mu\nu;\alpha\beta} - P_2^{\mu\nu;\alpha\beta}\right) \;.$$

Therefore, for a light-like vector $z$, one has

$$\Pi_{\langle\bar{q}Gq\rangle} = \Pi^{\mu\nu;\alpha\beta}_{\langle\bar{q}Gq\rangle}g_{\mu\alpha}z_\nu z_\beta = c(q^2)\frac{2(q\cdot z)^2}{q^2} \;.$$

This quantity reduces to the linear combination of $\langle\bar{q}Gq\rangle$ condensate contributions to the correlator for vector currents (see [7,3]). In this way, we get the formula

$$\Pi_{\langle\bar{q}Gq\rangle} = \frac{-320(q\cdot z)^2}{81q^6}\pi\alpha_s\langle\bar{q}q\rangle^2$$

from which we then obtain the fraction 80/81 appearing in Eqs. (14)–(15).

If $\mathbf{SD}_{\mu\nu;\alpha\beta}(q) = 0$, then we would have the same SRs both for $\rho$- and $b_1$-mesons. We process this hypothetical SR within the standard approach without $\alpha_s$-corrections in the perturbative contribution and obtain the following values for the low-energy parameters of a hypothetical $\rho b_1$-meson in Anti-Dual Nature:

$$m_{\rho b_1} = (0.865 \pm 0.030)\text{ GeV}\;;\quad f_{\rho b_1} = (0.162 \pm 0.005)\text{ GeV}\;;\quad s_{\rho b_1} = 1.58\text{ GeV}^2 \;. \tag{31}$$

We see that the mass and the decay constant of the $\rho$-meson are not so much affected by this (anti-)duality breakdown (10% for the mass). The case of the $b_1$-meson is quite opposite. Here the mass falls down to 45% (the decay constant, to 16%, see (21)). This seems to be quite natural. In the case of the $\rho$-meson, the deformation of the SR is large (the quark condensate contribution is enhanced by a factor of 3.25), but its functional dependence on the Borel parameter $M^2$ is almost the same. This is not the case for the $b_1$-meson. The deformation of the SR due to the opposite sign of the quark condensate contribution is essential and this results in such a large effect for the mass of the $b_1$-meson.

## QCD VACUUM TENSOR SUSCEPTIBILITY

The QCD vacuum tensor susceptibility $\chi$ has been introduced in [5,6] in order to analyze, in the QCD SR approach, the nucleon tensor charges $g_T^u$ and $g_T^d$. It is defined through the correlator (1)

$$\chi = \frac{\Pi_\chi(0)}{6\langle \bar{q}q \rangle} , \quad \Pi_\chi(q^2) \equiv \Pi^{\mu\nu;\mu\nu}(q^2) . \tag{32}$$

He and Ji [5] obtained for $\Pi_\chi(0)$ the value

$$\frac{1}{12}\Pi_\chi(0) \approx 0.002 \text{ GeV}^2 . \tag{33}$$

In order to obtain a reliable estimate in our approach we substitute the decomposition (2) in (32), use the relation (30) and arrive at the expression

$$\Pi_\chi(q^2) = 3\left(\Pi_+(q^2) + \Pi_-(q^2)\right) = 6\,\Pi^{\text{SD}}(q^2) . \tag{34}$$

This relation clearly demonstrates that $\Pi_\chi$ is formed by the **SD** part of OPE, *i.e.*, by the **four-quark condensate** contribution. Using the dispersion relations (6)

$$\frac{1}{12}\Pi_\chi(0) = \frac{1}{4\pi} \int_0^\infty \frac{\rho_+^{\text{phen}}(s) + \rho_-^{\text{phen}}(s)}{s} ds , \tag{35}$$

and the phenomenological models for spectral densities $\rho_\pm(s)$ in (8), the value of $\Pi_\chi(0)$ can be expressed in terms of mesonic static characteristics (the analogous formula has been published in [14]),

$$\frac{1}{12}\Pi_\chi(0) = \frac{(f_{b_1}^T)^2 - (f_\rho^T)^2 - (f_{\rho'}^T)^2}{2} + \frac{s_{\rho,\rho'} - s_{b_1}}{16\pi^2} = \begin{cases} -0.0055 \pm 0.0008 \text{ GeV}^2 & \text{[NLC]} \\ -0.0053 \pm 0.0021 \text{ GeV}^2 & \text{[Standard]} \end{cases}, \tag{36}$$

presented in (18)–(19) for NLC SR and (20)–(21) for the standard SR respectively[5]. Note here, both the resuts are very close one to another due to strong cancellations in the difference (36). So, just this combination accumulates the effect of the four-quark part of the whole condensate contribution. If we return to the example of an anti-dual Nature (see the end of the previous section, (31)) where this contribution is absent, we obtain the exact cancellation in (36), *i.e.* $\Pi_\chi(0) = 0$.

B&O in [2] have used the specific representation that leads to the decomposition

$$\Pi_\chi(q^2) = 12\,\Pi_1(q^2) + 6q^2\Pi_2(q^2) , \tag{37}$$

where $\Pi_1(q^2) = \frac{1}{2}\Pi_+(q^2)$ and $q^2\Pi_2(q^2) = \frac{1}{2}\left(\Pi_-(q^2) - \Pi_+(q^2)\right)$. Erroneously suggesting that $\lim_{q^2 \to 0}\left[q^2\Pi_2(q^2)\right] = 0$ and using the trick suggested in [13] based on the dispersion relation[6]

$$\Pi_\chi^{\text{n.p.}}(0) \equiv \frac{1}{12}\Pi_\chi(0) = \frac{1}{\pi}\int_0^\infty \frac{\rho^{\text{phen}}(s) - \rho^{\text{pert}}(s)}{s}ds , \tag{38}$$

they concluded that

$$\Pi_\chi(0)^{\text{n.p.}} = \Pi_1^{\text{n.p.}}(0) = (f_{b_1}^T)^2 - \frac{s_{b_1}}{8\pi^2} \approx -0.008 \text{ GeV}^2 . \tag{39}$$

---

[5] Depicted errors are obtained by a special invented $\chi^2$-criterium and take into account only the SR stabiliy.

[6] Here $\rho^{\text{phen}}(s)$ and $\rho^{\text{pert}}(s)$ are the corresponding spectral densities; the difference of these functions validates the usage of unsubtracted dispersion relation

But we see from our analysis that the value of $\left(\Pi_-(q^2) - \Pi_+(q^2)\right)$ is identically equal to 0 only in an absolutely self-dual world, which is definitely not realized in QCD

$$\frac{\Pi_-(0) - \Pi_+(0)}{2} = \frac{s_{\rho,\rho'} + s_{b_1}}{8\pi^2} - (f_\rho^T)^2 - (f_{\rho'}^T)^2 - (f_{b_1}^T)^2 = -0.0060 \pm 0.0017 \text{ GeV}^2 \text{ [NLC]}. \tag{40}$$

This value is comparable with the value of the B&O estimate (39) for $\Pi_1^{\text{n.p.}}(0)$ and should be definitely taken into account. Comparing two estimates, our (36) and B&O (39), one sees not so large deviation from one another. It should not be taken by surprise because radiative corrections significantly reduce the B&O value to $\Pi_1^{\text{n.p.}}(0) \approx -0.003$ GeV$^2$. For this reason the actual magnitude of our total correction to this estimate is of an order of 100%. When our paper was finished we find the paper [14] which contains an estimate of the correlator, $\frac{1}{12}\Pi_\chi(0) = -(0.0083 - 0.0104)$ GeV$^2$, using the constituent quark model. The authors of the paper have determined also a rather wide window for VTS by analog of Eq.(36) using QCD SR results from different literature sources: $\frac{1}{12}\Pi_\chi(0) = -(0.0042 - 0.0104)$ GeV$^2$. As we pointed out in the Introduction, since these meson constants appear in VTS in a form of a difference, one has to define them more precisely and in the framework of a unified approach. So the large width of this window is not surprise for us.

Finally, let us briefly discuss the effect of our estimate of VTS on the nucleon tensor charges. Here we follow to pioneering paper by He and Ji [6] where these charges were roughly estimated using two types of SRs. Our result (36) increases the lower (decreases the upper) boundary for the $g_T^u$ ($g_T^d$) charge approximately by a factor of 1.4:

$$g_T^u = 1.47 \pm 0.76 ; \tag{41}$$

$$g_T^d = 0.025 \pm 0.008 . \tag{42}$$

(The results of He and Ji $g_T^u = 1.33 \pm 0.53$ and $g_T^d = 0.04 \pm 0.02$ have been obtained for too low, in our opinion, value of $\Lambda_{\text{QCD}} = 100$MeV. We, instead, use the value of $\Lambda_{\text{QCD}} = 250$MeV.)

### Acknowledgments

This work was supported in part by the COSY Forschungsprojeht Jülich/Bochum. We are grateful to O. V. Teryaev, discussions with whom inspired this work, to R. Ruskov and N. G. Stefanis for fruitful discussions. One of us (A.B.) is indebted to Prof. K. Goeke and N. G. Stefanis for warm hospitality at Bochum University.

# Appendix

### APPENDIX A: EXPRESSIONS FOR NONLOCAL CONTRIBUTIONS TO SR

The form of contributions of NLCs to OPE (11) depends on a model of NLC. At the same time the final results of SR processing demonstrate stability to the variations of the NLC model provided the scale of the average vacuum quark virtuality $\lambda_q^2$ is fixed. Here we use the model (delta-ansatz) suggested in [7] and used extensively in [3]; the model leads to Gaussian decay for scalar quark condencate [7], $\langle \bar{q}(0) E(0,z) q(z) \rangle \sim \langle \bar{q}q \rangle \exp\left(-|z^2|\lambda_q^2/8\right)$ (see details in [7]), dominated in $b_\pm$ via $b_4$. Here the factorization hypotheses is applied for the four-quark condensate reducing its contribution to a pair of scalar condensates. In the NLC approach this may leads to an overestimate of the four-quark condensate contribution due to an evident neglecting of a correlation between these scalar condensates, see Fig.2.

In this model we obtain the "coefficients" for OPE in the SR (16)–(17)

$$\frac{b_\mp(M^2)}{\mp 2} = b_2(M^2) + b_3(M^2) \pm b_4(M^2) ,\qquad(A.1)$$

where $b_2$ corresponds to the vector ($\langle \bar{q}\gamma_m q \rangle$), $b_3$ to the quark-gluon-quark ($\langle \bar{q} G_{\mu\nu} q \rangle$) and $b_4$ to the four-quark ($\langle \bar{q}q\bar{q}q \rangle$) vacuum condensate contributions (here $\Delta \equiv \lambda_q^2/(2M^2)$),

$$b_2(M^2) = -16 ;\qquad(A.2)$$

$$b_3(M^2) = \frac{4\left(60 - 273\Delta + 359\Delta^2 - 134\Delta^3\right)}{3(1-\Delta)^3} ;\qquad(A.3)$$

$$b_4(M^2) = 24(\Delta - 7)\frac{\log(1-\Delta)}{\Delta} + 4\frac{25\Delta^2 - 21\Delta - 6}{(1-\Delta)^2} .\qquad(A.4)$$

The gluonic contribution $a_\pm$ coincides in this model with the standard expression (13). For quark and gluon condensates we use the standard estimates (for "renorm-invariant" quantities in (A.5) we do not refer to any normalization point)

$$\langle \frac{\alpha_s}{\pi} G^2 \rangle = 1.2 \cdot 10^{-2} \text{ GeV}^4 , \quad \langle \sqrt{\alpha_s}\bar{q}q \rangle^2 = 1.83 \cdot 10^{-4} \text{ GeV}^6 ,\qquad(A.5)$$

$$\lambda_q^2\left(\mu^2 \approx 1 \text{ GeV}^2\right) \equiv \frac{\langle \bar{q}\nabla^2 q \rangle}{\langle \bar{q}q \rangle} = \frac{\langle \bar{q}\left(ig\sigma_{\mu\nu}G^{\mu\nu}\right)q \rangle}{2\langle \bar{q}q \rangle} = 0.4 \pm 0.1 \text{ GeV}^2 .\qquad(A.6)$$

### APPENDIX B: INPUT PARAMETERS IN B&B AND B&O PAPERS

Both groups of the authors of [8] and [2] used different definitions of initial parameters for processing the SRs. Namely, B&O used the following set of values (without any indication on the scale at which renormalization non-invariant quantities are determined):

$$\alpha_s \approx 0.1\pi = 0.314 , \quad 4\pi^2 \langle \bar{q}q \rangle = -0.55 \text{ GeV}^3 , \quad \langle \sqrt{\alpha_s}\bar{q}q \rangle^2 \approx 0.61 \cdot 10^{-4} \text{ GeV}^6 ,\qquad(B.1)$$

whereas B&B[8] (on the scale $\mu^2 \approx 1 \text{ GeV}^2$)

$$\alpha_s = 0.56, \quad \langle \bar{q}q \rangle = (-0.250)^3 \text{ GeV}^3 , \quad \langle \sqrt{\alpha_s}\bar{q}q \rangle^2 \approx 1.37 \cdot 10^{-4} \text{ GeV}^6 .\qquad(B.2)$$

This resulted in approximately the same overall contributions of the quark condensate in both papers, see the end of Sect. 2.

---

[7] Here $E(0,z) = P \exp(i \int_0^z dt_\mu A_\mu^a(t)\tau_a)$ is the Schwinger phase factor required for gauge invariance.

[8] Let us remind that the standard value is $\langle \sqrt{\alpha_s}\bar{q}q \rangle^2 \approx 1.83 \cdot 10^{-4} \text{ GeV}^6$ [9].

[1] J. Govaerts, L. J. Rubinstein, F. De Viron and J. Weyers, *Nucl. Phys.* **B283** (1987) 706.

[2] V. M. Belyaev and A. Oganesyan, *Phys. Lett.* **B395** (1997) 307.

[3] A. P. Bakulev and S. V. Mikhailov, *Phys. Lett.* **B436** (1998) 351.

[4] Particle Data Group Booklet, July 1998.

[5] Hanxim He and Xiangdong Ji, *Phys. Rev.* **D52** (1995) 2960.

[6] Hanxim He and Xiangdong Ji, *Phys. Rev.* **D54** (1996) 6897.

[7] S. V. Mikhailov and A. V. Radyushkin, *Phys. Rev.* **D45** (1992) 1754; *Sov.J.Nucl.Phys* **49** (1989) 494; A. P. Bakulev and A. V. Radyushkin, *Phys. Lett.* **B271**, 223 (1991)

[8] Patricia Ball and V. M. Braun, *Phys. Rev.* **D54** (1996) 2182.

[9] M. A. Shifman, A. I. Vainshtein and V. I. Zakharov, *Nucl. Phys.* **B147** (1979) 385, 448.

[10] V. M. Belyaev, and B. L. Ioffe, ZhETF **83**, 876 (1982); A. A. Ovchinnikov, and A. A. Pivovarov, Yad. Fiz. **48**, 1135 (1988); A. A. Pivovarov, Kratk.Soobshch.Fiz. (Bull.Lebedev Phys.Inst.) $\mathcal{N}^{\underline{o}}$**5**, 3 (1991)

[11] M. D'Elia, A. Di Giacomo, E. Meggiolaro, *Phys. Rev.* **D59** (1999) 054503; E. Meggiolaro, hep-lat/9909068.

[12] D. Becirevic at. al., hep-lat/9808187; hep-lat/9808129.

[13] V. M. Belyaev and I. I. Kogan, *Int. J. Mod. Phys.* **A8** (1993) 153.

[14] Wojciech Broniowski, Maxim Polyakov, Hyun-Chul Kim and Klaus Goeke, *Phys. Lett.* **B438** (1998) 242.